\documentstyle[12pt]{article}
\begin{document}
\title{ \bf Lax Pair Covariance and Integrability of  Compatibility Condition }

\author{ S. B. Leble,\\
\small  
Faculty of Applied Physics and Mathematics\\
\small Technical University of Gdan'sk,  \\
\small  ul. G.Narutowicza, 11/12 80-952, Gdan'sk-Wrzeszcz, Poland,\\
\small   email leble@mifgate.pg.gda.pl\\
\small and\\
\small Kaliningrad State University, Theoretical Physics Department,\\
\small Al.Nevsky st.,14, 236041 Kaliningrad, Russia.\\}

\date {}
\maketitle
\begin{abstract}
We continue to study  Lax (L-A, U-V) pairs (LP) joint covariance with respect to Darboux
transformations (DT) as a classification principle.
The scheme is based on a comparison of general expressions for the transformed
coefficients of LP and its Frechet derivative. We use the compact expressions of the DT 
via some
version of nonabelian  Bell
polynomials. It is shown that one more version of Bell polynomials,
so-called binary  ones, 
form a convenient basis for
the invariant subspaces specification. Some non-autonomous generalization of KdV 
and Boussinesq equations  are
discussed in the context.
  Trying to pick up restrictions at minimal 
operator level we consider      
Zakharov-Shabat - like problem . The subclasses that allow a  DT symmetry 
(covariance at the LP
level ) are considered from a point of view of dressing chain equations. The case of
classic DT and binary combinations of elementary DT are considered with possible reduction
constraints of Mikhailov type (generated by  an automorphism). Examples
of Liuville - von Neumann equation for density matrix are referred as
illustrations .  

\end{abstract}

\thispagestyle{empty}

\section{Introduction}

If a pair of linear problems is simultaneously covariant with respect to a DT, it generates 
Backlund transformations of the corresponding compatibility condition. In the context
of such integrability the joint covariance principle \cite{L} may be considered as a
classification scheme origin. In this paper we examine realizations of such scheme looking for 
possible covariant form and appropriate basis with a simplest transformation properties.
It is important to note that the proof of the covariance theorems for the linear operators 
incoporates the so-called (generalized) Miura transformation having the form of (generalized)
Riccati equation. We give and examine here the explicit form of the equality in both general 
and stationary case,
look \cite{ZL} as well.
It gives additional nonlinear equation that is automatically solved by DT theory and used
for generation of t-chain equations \cite{BZ}.
We show how the form of the covariant operator may be found while some kind of Frechet derivatives
of the operator coefficients and the transforms are compared.

Further, it is clear that the choice of DT type defines the class of covariant operators. 
The "classic" DT generalization for the polynomials of a differentiation operator results in 
transformation formulas first obtained in Matveev papers \cite{M}, the form of DT we use below is
from \cite{ZL}.
It is also possible to do it in an abstract way: i.e. to 
define elementary DT (eDT) for any projector (idempotent) 
element in a differential ring or module \cite{ZL}. The abstract differentiation is 
naturally included in the scheme \cite{L}. The appropriate Zakharov - Shabat 
(ZS) problem which contains some elements as "potentials" and "wave functions" 
(WF) is studied. The results open the way for DT theory applications in 
arbitrary matrix dimension (including infinite - dimensional or strictly 
operator case). It is shown  \cite{U} that the sequence of $n$ such 
transformations in $n\times n$ matrix case with usual differentiation by a parameter
give the standard dressing formulas of ZS problem \cite{NMPZ} .
  
 In refs. \cite{LU}, \cite{U}  eDT for 
direct and conjugate problems were used to obtain 
the binary DT (bDT) for matrices and in   \cite{Le} 
- for
three projectors with applications to N-wave interaction,
a step from  \cite{LZ}  upto the general case \cite{Leb}.
The symmetric form of the resulting 
expressions for potentials and wave functions (WF) make almost obvious the 
heredity of reduction restrictions \cite{LU} and underlying automorphisms \cite{Mih}
of a generic ZS problems. In  \cite{LC} an application to some operator 
problem (Liuville - von Neumann equation) is studied.  

In the Sec. 2 we examine the general non-Abelian version of the operator transformations
constructed by means of the above-mentioned Bell polynomials generalization. The resulting
transformation formulas do not contain double summation that is convenient for the 
further analysis as in the context of classification as on the way of dressing chain 
reformulations \cite{W}, \cite{Sha}, \cite{VSha} and solution generation. 
In the last part of the Sec. 2 we list few such polynomials.
We show then the equation for the intermediate ($sigma$) function that naturally appear during
the Darboux theorem
proof. This equation generalizes Riccati equation (Miura transformation) for the celebrated KdV
theory for general dynamics - it may be called  generalized  Burgers equation.

 In the next Sec 3 we concentrate our attention on the algorithmic derivation of Lax pairs 
 starting from the idea of  covariance. We use  more simple nonabelian case to show main
 features of the scheme. The point we demonstrate is based on a comparison of the formal
 Taylor expansion of the transformed operator (by means of Frechet derivative notion) 
 of a Lax pair and its formal DT expression. 
 The comparison
 gives differential equations which solutions produce the explicit expressions for the Lax pair
 coefficients.
 
 Sec 4 introduces binary Bell polynomials and study the transformation of them via the 
 differential forms incorporated. We also prove invariance of some combinations of such
 BBP with respect to DT.

Next section starts from covariance of generalized ZS equation. Two versions of dressing chain
are constructed for stationary spectral problems with nonabelian spectral parameter. We are trying to 
pick up relations that do not depend on matrix representation \cite{ZS}. Simple examples are 
studied by closures of the chain equations on the operator level.
 
 The section 6 is devoted to the nonabelian Lax pairs from the point of view of bDT covariance. 
 The Abelian version of binary transform (commuting "matrix 
 elements") reproduces results of the dressing method based on the matrix 
 Riemann-Hilbert problem \cite{NMPZ}. 

 Here we mainly treat the dressing chain equations, sending readers to \cite{LCUK}, 
 where the equation
 $$
 -iX_t=[X,h(X)], \eqno(1.1)  
 $$
h(X) - analytical function, is studied. Further generalizations for essentially nonabelian
functions (e.g.  $h(X) = XA+AX, [A,x] \neq 0$, \cite{LC}) are considered 
 in \cite{CU},
 where abundant set of integrable equations is listed. The list is in a partial correspondence with 
 \cite{MS},
 and give direct link to solutions via the dressing chains. The papers also contain examples of 
 "self-scattering solutions" with discussions of possible applications \cite{CKLN}.

\section
{Darboux transformations in terms of Bell
polynomials.}

The solution of the problem of a polynomial linear differential
operator left and right division produces
a version of the Darboux theorem.  The theorem
was reformulated for the classical DT 
$$
\psi[1] = D\psi - \sigma \psi,  
\eqno(2.1)
$$ 
in terms of some generalization
of the  BeIl polynomials \cite{ZL} that give 
compact expressions for  transformed
coefficients of the operator. We reproduce here
the results with links to further applications, namely with 
equations for the element "$\sigma$" that appear inside the theorem proof.
 
Let us start from the linear operator
$$
L  = \sum_{n=0}^N a_n D^n, 
$$
 and the evolution equation (flow)
$$
\psi_t = L\psi. \eqno(2.2)
$$
Here the operator D may be a differentiation by some
variable and $\psi_t$ is the derivative with respect
to another one (see \cite{LZ} for generalizations).  
The transformation of the solutions of the
equation is taken in the standard form (2.1),
where 
$
\sigma = (D\phi)\phi^{-1}  
$
with a different solution of (2.2) incorporated
$
\phi_t = L\phi. 
$
We would now promote the convenient formulation of Matveev theorem \cite{M}

{\bf Theorem.}
{\it The  coefficients of the
resulting operator
$$
L[1] = \sum_{n=0}^N a_n[1] D^n
$$
are defined by
$$
a_N[1] = a_N,
$$
and for all other n, by
$$
a_n[1] = a_n + \sum_{k=n+1}^N [a_k B_{k,k-n} + 
((Da_k) - \sigma a_k) B_{k-1,k-1-n}] \eqno(2.3)
$$
that yields a covariance principle.}

It means that
the function $\psi[1]$ is a solution of the
equation
$$
\psi_t[1] = L[1]\psi[1],  
$$
where L[1] has the same structure and order as L.
The proof of this statement incorporates the equation that links $\sigma$ 
and coefficients of the operator L. It may be derived inside a factorization
theory \cite{ZL} and will be useful further
$$
\sigma_t = Dr + [\sigma,r], \qquad  r=\sum_{k=1}^N a_k B_k,\eqno(2.4)
$$
where $B_k$ are nonabelian Bell polynomials (see, e.g. \cite{SR}).
In the stationary case one has
$$
\sigma_t = \phi_{xt} \phi^{-1} - \phi_x \phi^{-1}\phi_t
\phi^{-1} = 0. \eqno(2.5)
$$
The functions $B_{m,n}$ are introduced in \cite{ZL}.
  
  We reproduce here the definition and some
statements about them.

{\bf  Definition }
$$
B_{n,\,0}(\sigma) = 1,\quad n=0,1,2,\dots,
$$
{\it and recurrence relations}
$$
B_{n,\,k}(\sigma)=B_{n-1,\,k}(\sigma)+DB_{n-1,\,k-1}(\sigma),\quad
k=\overline{1,\,n-1},\quad n=2,3,\dots.\eqno(2.6)
$$
$$
B_{n,\,n}(\sigma)=DB_{n-1,\,n-1}(\sigma)+B_n(\sigma),\quad
n=1,2,\dots.
$$
{\it define the generalized Bell polynomials} $B_{m,n}$.

The following  formula  is extracted,
$$B_{n,\,n-k+1}(\sigma)=\sum_{i=k}^n\,{i \choose
k}\,B_{n,\,n-i}(\sigma)\, D^{i-k}\sigma,\quad
k=\overline{1,\,n},\enspace n=0,1,2,\dots;
$$
$$B_{n+1}(\sigma)=\sum_{i=0}^n\,B_{n,\,n-i}(\sigma)\,D^i\sigma,\quad
n=0,1,2,\dots \eqno(2.7).
$$
\medskip
it gives the link between standard
(nonabelian) Bell polynomials and the generalized
ones:
$$
B_{n+1}(\sigma)=\sum_{i=0}^n\,B_{n,\,i}(v)\,D^{n-i}\sigma,\quad
n=0,1,2,\dots.
$$
Evaluation of the first three generalized Bell polynomials by the definition
gives
$$B_{n,\,1}(\sigma)=\sigma;\quad
B_{n,\,2}(\sigma)=\sigma^2+n\,D\sigma;\quad
B_{n,\,3}(\sigma)=\sigma^3+n\,\sigma'\,\sigma
 + (n-1)\,\sigma\,D\sigma+{n
\choose 2}\,D^2\sigma;$$

\section
{Towards the classification scheme. Joint
covariance of L-A pairs} 

The basis of the formalism we introduce is
elaborated starting from \cite{L} and  
the compact formulas with the generalized differential (Bell)
polynomials from the previous section.
  Note again, that it is valid for nonabelian
entries as well, the coefficients
$a_n$, solutions of the equation (1)
$\phi$ and $\psi$ we consider as  matrices or operators.
Let us however start from the scalar case.

To prepare the explicit
expressions for such work and show details we
would setup a couple of examples of the
theory.  Let us deliver a very simple
analysis for better understanding of the sense
of the integrability notion we introduce. First of all we notice that the "elder"
coefficients, with n = N and n = N-1 are
transformed almost trivially. It follows that in general the
functions rather do not  play the role of
potentials or unknown functions for a nonlinear
equation of a compatibility condition.  

If N=2, the general transforms (2.3) reduces to
$$
a_2[1] = a_2 = a(x,t),
$$
$$
a_1[1] = a_1(x,t) + Da(x,t) 
$$
$$
a_0[1] = a_0 + Da_1(x,t)+ 2a(x,t)D\sigma +  \sigma Da(x,t)\eqno(3.1)
$$
Keep in mind that we touch hear only the scalar (abelian,
more precise) case.  One can see that the explicit form  of the 
transformations really shows a difference between the 
coefficients $a(x,t),a_1(x,t)$  that transform  without solutions account 
and the $a_0=u(x,t)$ to be an unknown
function of
a forthcoming nonlinear equation that we call potential in
the context of Lax (L-A-pair) representation. One may
easily recognize the KdV case here. Namely when
$a = const, a_1 = 0$ , $a_0$ plays   the role
of the only unknown function of the KdV
equation. 
So we may formulate an 

{\bf observation:} {\it The abelian
case N = 2 is the first nontrivial example of 
covariant operators set with coefficients
$a_{1,2}$ that depends only on x and 
additional parameter (say t) but its transforms contain the only functions 
$a_{1,2}$, hence, to be called trivial. The transformation (DT) for u is given by the last equation
of (3.1) and depends  on both $a_{1,2}$ and solutions of the equation (2.2) via $\sigma$.}

By the next order example we would show more details
especially  
when  pairs of operators are analysed simultaneously. Especially it is 
important when both operators are functions of the only potential. 
Let us take N = 3.
$$
b_3[1] = b_3,
$$
$$
b_2[1] = b_2 + Db_3' 
$$
$$
b_1[1] = b_1 + Db_2 + 3b_3 D\sigma  +  \sigma Db_3    
$$
$$
b_0[1] = b_0 + Db_1 +  \sigma D b_2 + (\sigma^2 + (2D\sigma))Db_3 +3b_3 (\sigma D\sigma +
D^2\sigma) \eqno(3.2)
$$

Consider (3.1) and (3.2) as coefficients of a Lax pair
operators both depending on the only  variable
u  simultaneously and suppose the coefficients
of the operator are analytical functions of u and
its derivatives with respect to x.
Now we change $t \rightarrow y$ and $L \rightarrow L_1$ in the equation (2.2)
corresponding to the case (3.1) and left the parameter t ($L \rightarrow L_2$) for the case (2.5),
forming a Lax pair
$$
\psi_y = L_1\psi.  \eqno(3.3)
$$
$$
\psi_t = L_2\psi. \eqno(3.4)
$$
To produce the KdV case generalization we go to a stationary in y
solution of (3.3). 
$$
\phi_y = \lambda \phi,
$$
where $ \lambda $ is a constant.

 Let us also recall the KdV case. Then the stationary version of (2.4) for N = 2 is
$$
\sum_0^2 a_nB_n = c,
$$
that
reads as
$$
\sigma^2 + \sigma' + u = c = const.\eqno(3.5)
$$
Note 
that the  equation (2.5) for N = 3 is still valid for
this $\sigma = \phi_x \phi^{-1}$, if  $\phi$ is
a solution to the Lax pair system (3.3,3.4).
If in the equations of (3.2) we restrict ourselves by
the case of $b_2 = 0$ and $ b_3 = b = const$, we arrive at the second  equation of the 
KdV Lax pair.

Next, returning to the general case and  taking into account the triviality of 
$b_3=b(x,t)$ and $ b_2$ in the above-mentioned sense, the first non-trivial potential is 
$$
b_1 = F(u,u',...). \eqno(3.6)
$$
 Suppose
further that the covariance principle is valid,
or, in shorten arguments notations, we address to
an equation for this function F.
$$
b_1[1] = F(u[1]) = F(u + Da_1 + 2a D\sigma +  \sigma Da  )  
= F(u) +  Db_2 + 3b D\sigma  + \sigma Db. \eqno(3.7)
$$
Analyticity of F 
gives a  possibility to use Taylor series  
expansion in the left side of (3.7)
$$
 F(u[1]) = F(u) +   F_u (2a D\sigma+ Da_1 +\sigma Da )+ F_{Du}(...)+... . \eqno(3.8)
$$
In fact we compare the transform (3.7) with the Frechet differential (3.8) of the F.
The  equation holds identically if the coefficients by the $\sigma, D\sigma$ and the free term 
are the same.
Introducing $F_u=c(x,t)$ yields
$$
 2 a c = 3 b,\eqno(3.9)
$$
or 
$$
F(u) = \frac{3bu}{2a}  . \eqno(3.10)
$$
with additional conditions
$$
cDa_1=Db ;  \eqno(3.11)
$$
$$
cDa=Db \eqno(3.12)
$$
Plugging c from (3.9) into (3.12),  
one go either to 
$3 D(lna) = 2 D(lnb) $
and, integrating, obtains 
$$
b=a^{3/2}c_1(t), \eqno(3.13)
$$ 
or to $Da=Db=0$. In the last case (3.12) is valid with arbitrary c, or independent b(t) and a(t).
While (3.11) yields the equation for $a_1$ for both cases
$$
3  Da_1=  = 2a Db/3b \eqno(3.14)
$$
with arbitrary $c_1(t)$. 
 
Next
conditions follow from the last equation of the
system (3.2), i.e., if one introduces next
analytical function G and denote
$$
b_0 = G(u, u',...), \eqno(3.15)
$$
the transformed $b_0$ gives  
$$
G(u + Da_1 + 2a D\sigma +  \sigma Da  )  = G(u) +  
G_u (Da_1 + 2aD\sigma +  \sigma Da )
+  G_{Du}D(Da_1 + 2a D\sigma +  \sigma Da )+ ... .\eqno(3.16)
$$
The DT transformation formula for the potential u is obviously used.
The DT for the last coefficient $b_0$, see (3.2), yields
$$
b_0[1] = G(u) + Db_1 +  \sigma Db_2 + (\sigma^2 +  2(D\sigma))Db +
3bD(\sigma^2/2 + D\sigma).   \eqno(3.17)
$$
We should account now the general version of the Miura transformation (3.5) 
that has the form
$$
\sum_0^2 a_nB_n = u+a_1\sigma+a(\sigma^2+D\sigma )=\mu,
$$
by which we would express the $ \sigma^2 $ in (3.17).
Doing this and equalizing the expressions  (3.16) and (3.17) yields
$$
D\frac{3bu}{2a} +  \sigma Db_2 + ((\mu - u - a_1\sigma)/a  + (D\sigma))Db +
3b D [(\mu -u-a_1\sigma)/2a 
+(D\sigma)]=
$$
$$
G_u (Da_1 + 2a D\sigma +  \sigma Da
+  G_{Du}D(Da_1 + 2a D\sigma +  \sigma Da )\eqno(3.18)
$$
The equation (3.18) gives for the coefficients: 
by $D^2\sigma$
$$
G_(Du)2a =3b,\eqno(3.19)
$$
by $D\sigma$, taking (3.19) into account
$$
G_u 2a +9b(Da) /2a = Db-3ba_1/2a  ,\eqno(3.20) 
$$
by  $\sigma$
$$
G_u Da + \frac{3b}{2a}D^2a(x,t) = Db_2-a_1/a-3b D(a_1/2a), \eqno(3.21) 
$$
and the free term is
$$
D \frac{3bu}{2a}+((\mu -u)/a)Db + 3b D [(\mu -u )/2a]=
G_u Da_1 +  3b(D^2a_1)/2a. \eqno(3.22) 
$$
From   (3.19) and (3.20) we express $G_u$:
$$ 
 G_u = Db/2a-3ba_1/4a^2-9b(Da) /4a^2.\eqno(3.23)
 $$
If  $G_u$ is nonzero, from (3.21) follows
$$
 (Db/2a-3ba_1/4a^2-9b(Da)/4a^2 )Da+\frac{3b}{2a}D^2a  = Db_2-a_1/a-3b D(a_1/2a)
$$
The free term (3.22) gives
$$
u\frac{Db}{2a} +\mu (Db /a - \frac{3bDa}{2a^2} =
 (\frac{Db}{2a}-\frac{3ba_1}{4a^2}-\frac{9bDa}{4a^2})Da_1(x,t)+  3b(D^2a_1)/2a. \eqno(3.24) 
$$
When u is linear independent from $\sigma$ and derivatives, and we do not account higher terms in
thr Frechet differential, the only choice
$ Db=0$ kills the term with u, and  (3.24) simplifies
$$ 
 D^2a_1  -a_1(Da_1)/2a = 0. 
$$
The condition Da=0 as the corollary of (3.12) is taken into account.
The equation (3.21) also simplifies
$$Db_2-a_1/a-3b (Da_1)/2a=0
$$
and integration gives the expression for $b_2$.

Another possibility is $G_u=0$. It gives $9b(Da) /2a = Db-3ba_1/2a$ instead of (3.23). The
free term transforms as 
$$
u\frac{Db}{2a} +\mu (Db /a - 3bDa/2a^2)=
 +  3b(D^2a_1)/2a. \eqno(3.25) 
$$
and by the same reasons gives the conditions $ Db=Da=0$. It further means $a_1=0$ and, finally, 
from (3.21),
$$
Db_2 =0.
$$
Hence this case contain KdV equation with the possible $a(t), b(t),b_2(t)$.  

{\bf Remark 1.}  The results for one isolated equation (3.3) contain rather wide class of coefficients 
(in a comparison with the joint covariance of (3.3) and (3.4). Namely, the $a, a_1$ are arbitrary 
functions of x,t. It may be useful for
construction of potentials and solutions (e.g. - special functions) for the linear Schr{\"o}dinger and 
evolution equations of the one-dimentional quantum mechanics \cite{Y}. 
The KdV case may be described separately (denote further Df = f'):
$$
  G_u\sigma' +   G_{u'} \sigma'' =
 3b(1-a) u'/4a^2+3b 
\sigma''/4a 
$$
The only
choice is possible, if we consider $ \sigma, \sigma', u' $ as  
independent variables 
$$
  G_u = 0,
$$
$$  G_{u'} = 3b/4a
$$
or, taking into account the condition of zero coefficient by $u'$, $a = 1$, 
one arrives at
$$
G(u,u',...) = 3bu'/4.
$$ 
The result leads directly to one of equivalent Lax pairs for the KdV equation.

Now we would consider the next natural example with interchanged equations
 defining a spectral problem and evolution.
 Hence we should start from the equation (3.3) for the third order spectral operator,
 using (3.4) as evolution with N=2.
 
 We would restrict ourselves by
 the case of $b_2 = 0$ and the autonomous $ b_3 = 1$ ,  $b_1 = const =b$, $b_0 = u$.
 The DT yields
 $$
 b_1[1] = b_1 + 3b \sigma' \eqno(3.26)
 $$
 $$
 b_0[1] = b_0 + b_1' + 3b (\sigma_x \sigma +
 \sigma_{xx}). \eqno(3.27)
 $$
 
 As it was shown by analysis of the third order operator, see (3.21-25),
 the covariant spectral problem  has the form
 $$
 \psi_{xxx} + b \psi_{x} + G \psi  = \lambda \psi \eqno(3.28)
 $$
The second (evolution) equation of the case is :
 $$
 \psi_t = - \psi_{xx} - w \psi \eqno(3.29)
 $$ 
   If one consider (3.3)( specified in (3.26) and (3.27)) or (3.28,29) 
 as the Lax pair equations 
  both with coefficients depending on the only  variable
 w. Suppose again that the coefficients
 of the operators are analytical functions of w and
 its derivatives (or integrals) with respect to x. If one wants to save the form of 
 the standard DT for the variable w 
 (potential) the analysis just similar to the previous example give for the constant $\alpha$
 $$
 b  = 3w/2 + \alpha. \eqno(3.30)
 $$
 Compare with the formula (3.25), or $w[1] = w + 2\sigma_x$.
 Then the transformation for the potential w
  follows from the last equation of the
 system (3.27), i.e., 
 $$
 G[1] = G + 3w_x/2 +  3(\sigma^2/2 +
 \sigma_{x})_x. \eqno(3.31)
 $$
 we see that further analysis is necessary due to the possible constraint (reduction) existence.
 Then, again similar to the previous case of KdV 
 the covariant equations (3.27,28) are accompanied by the following (Burgers) 
 equation.
 $$
 \sigma_t = -( \sigma^2 + \sigma_x)_x - w_{x} \eqno(3.32)
 $$
 for the problem (3.28) and
 $$
 \sigma^3 + 3\sigma_x\sigma + \sigma_{xx} + b\sigma + u = const, \eqno(3.33)
 $$
 see (2.5), compare with (3.5) that was  Riccati  equation (stationary version) 
 for the second order spectral problem
 corresponding to KdV. If one would use the equation (3.32) in (3.31), the time-derivative of w 
 appear. Moreover, the further analysis shows that the case we study need to widen the 
 functional dependence in u, namely we should include not only derivatives of w with respect to x,
 but integrals (inverse derivatives) as arguments of the potential. So, for the equation (3.28) 
 let us  introduce next
 analytical function G and denote
 $$
G = G(\partial^{-1}w, w, w_x,...\partial^{-1}w_t,w_t,w_{tx},...), \eqno(3.34)
 $$
   the first terms of 
 Taylor series for (3.34) read 
 $$
 G(w + 2\sigma_x) = G(w) +  G_{w_x} 2\sigma_{xx} + G_{\partial^{-1}w_t} 2\sigma_t+ ... , \eqno(3.35)
 $$
 where we show only terms of further importance. The DT transform (24) after substitution of 
 (3.31) gives 
 $$
   G(w) + 3w_x/2 +  3(- \sigma_t - w_x)/2 +
 3\sigma_{xx}/2. \eqno(3.36)
 $$
 Equalizing (3.36) and (3.35) one obtains
 $$
 G_{w_x} = 3/4; G_{\partial^{-1}w_t} = - 3/4,
 $$
 That leads to the exact form of the Lax pair for the Boussinesq equation from [2] for the choice
 of $\alpha = -3/4$.

 {\bf Remark 2.} As one could see we cut the Frechet differential formulas on the level that is 
 necessary  for the minimal flows. The account of higher terms lead to higher 
 flows (higher KdV, for example).

{\bf Remark 3} In the second example we obviously restrict the description by the autonomous case 
within the special choice 
of DT and do not consider the class of gauge  equivalent equations. The general case with 
account of 
higher derivatives will be 
presented elsewere  in forthcoming publications.

We would finish with the 

{\bf Theorem.} {\it The Darboux covariant Lax pair (3.3,4) with the stationary equation (3.3) where the operator
$L_1$ is given by (3.1)  has the
following
coefficients:
$$
a_2=a(t),
$$
$$
a_1 = a_1(x,t),
$$
is a solution of the equation $Da_1 - a_1^2/4a = c_2(t)$ with arbitrary $c_2$,
$$
a_0 = u.
$$
The evolution (3.4), which form is presented by (3.2), contains
$$
b_3 = b(t),
$$
$$
b_2 = \int a_1 dx /a+3b a_1 /2a , 
$$
$$
b_1 = 3b/2a,
$$
$$
b_0 = -3ba_1u/4a^2 + 3b(Du)/2a.
$$
The pair produce the generalization of the KdV equation 
$$
L_{1t} = [L_1, L_2]
$$
that is solved by standard DT}.

\section
{  Binary Bell polynomials under Darboux
transformation} 

  The binary Bell polynomials (BBP) {\cal Y} that we are going to use are defined in terms of
the exponential differential polynomials:

$$Y_{mx,nt} (f) \equiv e^{-f} \partial_x^m \partial_t^n e^f \eqno(4.1)$$
\noindent by means of the link between $\cal Y$-polynomials and the standard
Hirota expressions \cite{GLNW}:

$$D_x^p D_t^q G' \cdot G \equiv \left(\partial_x - \partial_{x'}
\right)^p\left( \partial_t - \partial_{t'}\right)^q G'(x,t)
G(x',t')\raise-1.5mm\hbox{$\Big\vert$}_{x'=x,t'=t}. \eqno(4.2)$$
The connection is given by the identity:
$$
{\cal Y}_{mx,nt} (v , w) \equiv \exp[\frac{v-w}{2}]\exp[\frac{-v-w}{2}]
D^m_x D^n_t  \exp[\frac{v+w}{2}] \cdot   \exp[\frac{w-v}{2}] \eqno(4.3)$$
They inherit the easily recognizable partitional structure of the
Bell polynomials (BP):
$$ 
 {\cal Y}_x(v)=v_x, \quad {\cal Y}_{2x}(v,w)=w_{2x} + v^2_x,
\quad {\cal Y}_{x,t} (v,w)=w_{xt} + v_xv_t,  
 $$
 $${\cal Y}_{3x} (v,w) = v_{3x} + 3v_x w_{2x} + v^3_x,\ \cdots
\eqno(4.4)$$

The expression of the  BBP  contains the structure we studied in previous sections.
Therefore the covariance of the expressions can be examined in similar way.
  To check the conjecture
about  covariance of a BBP one should
confirm the coincidence of the expressions of
such polynomial generated by (4.3) and
transformations of functions (2.3) .  
Let, as the simplest example, consider the second order
polynomial from the point of view of preceding
analysis reproducing the addition formula from \cite{LS}. We
get
$$
{\cal Y}_{2x}(v, v + q) = P_0 B_2(v) +
P_{2x}B_0. 
$$
Here the symbols ${\it Y}_{2x}, P_{2x}(q)$ are
also defined in \cite{GLNW},\cite{LS}  and $B_n$ are usual BP.
$${\cal Y }_{2x}(v, v + q) = \psi^{-1}L \psi =
\psi^{-1}(\psi_{xx} + q_{2x}\psi)
$$
Thus, as it was shown in the Sec. 3, the
polynomial contains  D-covariant expression for
the operator  L (N=2) with the potential $q_{2x}$. It means that the spectral
problem for KdV case may be expressed by means of the BBP, or
$${\it Y}_{2x}(v, v + q) = \lambda.\eqno(4.5)$$
This simple example shows how to find the basis in
which it is convenient to make the
transformations. It turns out easier than the
direct test  and allows to forward the
business for arbitrary N.

The general expression for the BBP, transformed like (2.3) \cite{GLNW} is the addition formula
$$
{\cal Y}_{Nx}(v, v + q) =
\sum_{p=0}^{E(N/2)}{N \choose 2p}P_{2px}Y_{(N -
2p)x}(v)\eqno(4.6),
$$
where E(N/2) is the integer part of N/2.
After the logarithmic linearization (16),
$$
{ \cal Y}_{Nx}(v, v + q) = \sum_{m = N -
2E(N/2)}^{E(N/2)}{N
\choose N-m}P_{(N-m)x}(q)D^m\psi = \sum_{m=0}^N
a_m^PD^m \psi, \eqno(4.7)
$$
the expression now coincide by the form with the
right-hand side of a 
Lax equation with the operator L.
We see the coefficients of the DO in (4.7), which
DT is determined by   (2.3).
The extracted coefficients are
$$
a_m^P = {n \choose N - m}P_{(N-m)x}(q),
$$
if $N-m$ is even, otherwise $a_m^P = 0$ (for odd
$N-m$. Or in other words,   $a_m^P $   are
nonzero only in the set of $m \in \{N, N-2, ...,
N - 2E(n/2). $ Substituting the coefficients from
(4.7) with m = N - 2p into (2.3), one obtains
$$
a_m^P [1] = a_m^P + \sum_{j = m+1}^N\prime {N \choose (N-j} qP_{(N-j)x}(q)[B_{j,j-m} - \sigma
B_{j-1,j-1-m}]$$
$$ 
+ P_{(N-j)x}(q)'B_{j-1,j-1-m}\}.\eqno(4.8)
$$
The terms in the sum of (4.8) are nonzero also
if N - j = 2p is even, it is marked by the
prime in the sum. The definition, convenient
formulas  and explicit expressions to the
generalized Bell polyinomials $B_{n,m}$ are
given in the Sec.2.

For more illustration of the technique let us take
again the minimal polynomials and evaluate the
r.h.s of (4.7) and (4.8). First take an even N and
begin with
$$
a_{N-2}[1] = a_{N-2} + a_N(B_{N,2} - \sigma
B_{N-1,1}) =  a_{N-2} + N\sigma'. 
$$
We have taken into account that $a_N = P_0 = 1$
Compare with the addition formula   
$$
P_{2x}(q + 2v) = {2 \choose 0}P_{2x}(q) + {2
\choose 2} P_{2x}(2v) = q_{2x} + 2 v_{2x} =
q_{2x} + 2\sigma'.
$$
We see that in the case of N = 2 the polynomial
is invariant.

The example of odd polinomial N =3 yields
$$
{\cal Y}_{3x}(v,v+q) = {\cal Y}_{3x}(v) + 3q_{2x}v_x = 
\psi^{-1}(D^3 + 3q_{2x}D)\psi \eqno(4.9) 
$$
with the known
transform of the operator inside (4.9). There are
two possible scopes of an incorporation of the
polynomials into Lax pairs. Either it belongs to
evolution equation or results in a spectral
problem of third order and leads to SK or Boussinesq equations.

Let us first comment the KdV case, so the spectral
problem
$${\cal Y}_{2x}(v, v + q) = \lambda.\eqno(4.10)$$
introduces a potential $q_{2x}$ to be a dependent
variable for the KdV equation. We should rewrite
now the evolution equation in terms of binary Bell
polynomials. Starting from (4.10) one notice that
the corresponding coefficient of (4.10) are
defined by the equation (4.10):
$$
a_3^P = 1, \qquad a_1^P = 3q^{(3)}_{2x},
$$
$$
a_2^P =  
a_0^P = 0.
$$
The transforms of them is determined by (4.9) or
one may also use the explicit expressions from
(4.5). We should write, for example,
$$
3a^P_1[1] = 3a^P_1 + 3\sigma'.
$$
It follows that the potential (dependent variable)
should be choosen as 
$$q^{(3)}_{xx} = q_{xx}/2 \eqno(4.11)$$
 to fit
the transformation law for $q_{xx}$.
If one examine the transformation of the
coefficients of the general operator (e.g.(4.6)),
one notices that the "youngest" coefficient $b_0$
is changed by necessity. The addition depends on
a potential and $\sigma$. So the zero value of
$a_0^P$ could not be generally preserved. We
arrive to the

{\bf observation:} {\it the isolated ${\cal Y}_{3x}$ is not
generally covariant}.

Therefore there is a stright necessity to
incorporate different polynomials to support the
covariance. So we take the combination
$$c_3 {\cal Y}_{3x}(v,v+q^{(3)}) + c_2{\cal Y}_{2x}(v,v+q^{(2)}) + c_1
{\cal Y}_{1x} \eqno(4.12)$$ 
with different potentials $q_i$.
If we restrict our choice by the unique potential
that enter the spectral problem  (4.5) and note
that the choice of $q^{(3)} = q/2$ is fixed by (4.11),
we get the link of $q^{()}$ and $q$ from (4.5) that
gives
$$
q^{(2)} = q + 3c_3q_x/(4c_2)\eqno(4.13)
$$
it means that   
constants
$c_i$ are arbitrary and influence only the form
of the Lax pair and the compatibility condition.

{\bf Proposition.} {\it The combination (4.12) by the condition (4.13)
is covariant with respect to DT.
If $c_2 = c_3 = 1$, the compatibility condition of (4.5) - (4.12) is equivalent to the equation  
$$
u_t = 3uu_x/2 + u_{xxx}/4
$$
that is nothing but one of KdV forms.}

Many of other integrable equations may be incorporated 
into the scheme to be developped.

 \section{The example of nonabelian Zakharov-Shabat problem}
 
 Here we outline the general scheme of the dressing chain derivation in the nonabelian case.
 Let us start from the evolution
 $$
 a_{0}\Psi +a_{1}D\Psi =\Psi _{t}\eqno(5.1)
 $$
 that contain the simplest form of the polynomial L(D)-operator. The case
 is the nontrivial 
 example of a general equation (2.3) with non-abelian entries.
 We would respect $\Psi$ or other (necessary for DT construction) solution
$ \Phi$  as operators. The equation  enter to the Lax pair of 
 integrable nonlinear equations ( e.g. - Davey-Stewartson) while
 its stationary versions produce Nonlinear Schr\"odinger case.
 It is also interesting by itself: quantum one-dimensional Pauli or Dirac equations 
 after multiplication by the
 appropriate matrix takes the form of the equation (5.1).
 
 The potential $a_{0}$ may be
 expressed in the terms of $\sigma$ from the stationary version of the equation (2.4,5),
 when $\sigma_t = 0,$. Namely, introducing the iteration index i, we have the link
 $$
 Da_{0}^{i}+[a_{0}^{i},\sigma_{i}]=-Da_{1}-[a_{1}, \sigma_{i}]\sigma_{i}.\eqno(5.2)
 $$
 the connection is linear, but contains the commutator. Let us denote 
$ad_{\sigma}x=[\sigma,x]$. Then 
 $$
 a_{0}^{i}=(D-ad_{\sigma_{i}})^{-1}\{-Da_{1}-[a_{1},\sigma_{i}]\sigma_{i}.\}\eqno(5.3)
 $$
 The existence of the inverse operator in (5.3) need some restriction for the
 expression in \{,\} brackets, the expression should not belong to the kernel
 of the operator $D+ad_{s_{i}}$. In the subspace, where the Lie product is zero, 
 the equation (5.2) simplifies.  
 The DT is also simple. 
 $$
 a_{0}^{i+1}=a_{0}^{i}+[a_{1},\sigma_{i}]\eqno(5.4)
 $$
 Note that $a_{1}$
 is not transformed due to the general formulas (2.3). 
 Substituting the link (5.3) for i,i+1 into (5.4) one arrives to the chain
 equations. One also could express matrix elements of $a_{0}$ in terms of the
 elements of the matrix $\sigma$ and plug into the Darboux transform (5.4)
 separately. 
 
 Let us give more details of the construction in the stationary case. Let us denote $a_{0}=u$,
 $a_{1}=J$, choosing the case $DJ=0$ and $\Psi, \Phi$ correspond  $\lambda, \mu$.
 Note that there are two possibilities for stationary equations from nonabelian (5.1):
  either 
  $$ \Psi _{t} = \lambda \Psi 
  $$
  or
  $$ \Psi _{t} = \Psi \lambda \eqno(5.5)
  $$
  and the first of them  leads to the essentially   trivial connection between solutions
  and potentials from the point of view of DT theory \cite{MS}. Namely, if 
  $$
  J\Phi_x+u\Phi=\mu\Phi,
  $$
  then
  $$
    \sigma = \Phi_x\Phi^{-1}=J^{-1}(\mu-u)\eqno(5.6)
   $$
and DT do not contains eigen functions. It means, for example, that starting from 
constant potential one never obtain x-dependent u by iterations.
 
In the second case one writes
$$
\sigma = \Phi_x\Phi^{-1}=J^{-1}(\Phi\mu\Phi^{-1}-u)
$$
and DT takes the following form
$$
   u^{i+1}=u^{i} - J^{-1}[s^i,J]+J^{-1}[u^{i},J]=J^{-1}u^{i}J- J^{-1}[s^i,J],\eqno(5.7)
$$
where it is denoted $s = \Phi\mu\Phi^{-1}$, here and further iteration number indices omitted.
The potentials $u^i$ may be excluded from the equation (2.4) for this case
$$
\sigma_t=Dr+[r,\sigma], 
$$
with 
$$r=J\sigma+u=s\eqno(5.8)
$$.
The stationary case, after the plugging $u^i$ from (5.8) and returning indices, gives
$$
s^{i+1}=s^{i}+J \sigma^{i+1} -\sigma^{i}J \eqno(5.9)
$$
and the link of s with a $\sigma$  is very simple
$$
Ds^{i}+[s^{i},\sigma^{i}] \eqno(5.10)
$$
The formal transformation that leads to the chain equations is similar to the (5.3), 
namely, after substitution 
$$
\sigma^{i}=-ad_{s_i}^{-1}Ds^i
$$
 into the equation (5.9).
 
 Further progress in the development of this programme is connected with the choice 
 of the additional algebraic structure over the field we consider. It can be useful for 
 the concrete representation of solutions of the equation (5.10).
 For example, if the elements
 $s^{i},\sigma^{i}$ belong to a Lie algebra with structure constants 
 $C_{\alpha \beta}^\gamma$, then, after the choice of the basis $e_\alpha$ one 
 introduces the expansions
 $$
 s^{i}=\xi_\alpha^i e_\alpha
 $$
 and
 $$
  \sigma^{i}=\eta_\alpha^i e_\alpha.
 $$
 Plugging into (5.10) gives the differential equation
 $$
 D\xi_\alpha^i+C_{\gamma \beta}^\alpha\xi_\gamma ^i\eta_\beta^i=0.
 $$
 If one defines the matrix 
 $$B_{\beta\alpha}=C_{\gamma \beta}^\alpha\xi_\gamma ^i, \eqno(5.11) $$ 
 then, outside of the kernel of B
 $$
 \eta_\beta^i=-B^{-1}_{\beta\alpha}\xi_\alpha^i \prime.
 $$
 By the definition of the Cartan subalgebra C the correspondent subspace does not
 contribute in the Lie product of (5.10). 
 
 {\bf Statement.} {\it 
 If, further, J belongs to a module over 
 the Lie algebra, $Je_\alpha = J_{\beta\alpha}e_\beta$ and there exist an 
external involutive 
 automorphism such that $e_\alpha^+ = -e_\alpha,$ $ C_{\alpha \beta}^{\gamma*}=
C_{\alpha\beta}^{\gamma*}$, $e_\alpha J = J_{\alpha\beta}e_\beta$. Then, 
 the chain equation for the variables $\xi_\alpha^i $
 takes the form}
 $$
 \xi_\alpha^{i+1} = 
 \xi_\alpha^i  - B^{-1}_{\beta\gamma}\xi_\gamma^{i+1}\prime J_{\beta\alpha}+
 J_{\alpha\beta}B^{-1}_{\beta\gamma}\xi_\gamma^i\prime ,
 $$
 {\it where the matrix B is defined by (5.11) and the components 
$e_\alpha$ outside of C}. 
 {\it Otherwise 
 $$
 D\xi_\gamma^i=0,
$$
 if $e_\gamma \in C$}.
 
 The system of differential equations
 is hence nonlinear as the matrix B 
 depends on $\xi_\gamma ^i.$
 The related scheme is developed in \cite{Sh}.

 {\bf Remark.} The scheme may be generalized for  a nonstationary equation (2.5).
 The equation (5.2)
 then have the additional term $D\sigma^{i}_t$ from the r.h.s.

 \section{Covariance with respect to Binary DT and Dressing Chains}
 
 In this section, as in previous, we do not touch directly the joint
 covariance with respect to binary DT (bDT)and 
 classification problem,
 sending the reader
 to the publications \cite{LC},\cite{LCUK},\cite{CKLN}  an, especially to most general 
\cite{CU}, where the nobelium functions and rational dependence on parameter are admissible.
Those papers are devoted to evolution in quantum  statistics: the equation of Liuville-von Neumann
is considered as compatibility condition of stationary ZS problem algebraic reduction and 
equation that introduce a time variable.

 Hence we continue to study stationary version of the ZS problem (5.1)
$$
 JD\Psi+u\Psi =z \Psi \eqno(6.1)
$$
 and dressing chains of bDT (see the Introduction,  
 \cite{MS}, compare with  e.g. \cite{Leb}, the definition of which we use here). The algebraic 
 reduction imply $D\Psi = - \mu\Psi, D\mu =  0,\mu is a scalar. [$ After the reduction we use the
 straight corollary
 of (6.1), changing the notation as in \cite{LC}, $J\rightarrow A, u\rightarrow \rho$:
 $$
 \mu A\varphi + \rho\varphi  =z_\mu \varphi,\eqno(6.2)
 $$
 and the conjugate problem
 $$
  \nu\chi  A+\chi \rho    = z_\nu \chi,\eqno(6.3)
 $$
 Let the projector P have a representation
  $$
  P = \varphi (\chi, \varphi)_p^{-1} \chi,                         \eqno(6.4)
 $$
 for a given constant idempotent p \cite{Leb}. 
 The bDT formula connects the elements (operator valued potentials - 
 density matrix in the physical context),
 $$\rho[i+1]=\rho[i]+(\mu_i-\nu_i)[P_i,A]\eqno(6.5)
$$. 
 We would rewrite the conjugate spectral problems \cite{LC} in terms of P, by 
 means of the definition (6.4)
 $$(\rho - \mu A)P  = z_\mu P$$
 $$P(\rho  - \nu A) = z_\nu P$$
  Each equation links  the 
 "potential" $ \rho $ and
 the projector $P$ that plays the role of the element $\sigma$ of 
previous sections.
 $$ 
 \rho P = \mu A P+z_\mu P 
 $$ 
 $$
 P\rho  = \nu P A+z_\nu P. \eqno(6.6)
 $$
 The importance of both of them is obvious from the following corollaries  
 $$P(\rho-\mu A)P
 =(\rho-\mu A)P\label{PLP1a}\\
 $$
 $$
 P(\rho-\nu A)P
 =  P(\rho-\nu A).
 $$
 Let us also note that 
 $$
 (\nu-\mu)PAP = (z_{\mu}-z_{\nu})P \eqno(6.7)
 $$
 and
 $$
 P \rho P = \frac{\nu z_{\mu}-\mu z_{\nu}}{\nu-\mu} \eqno(6.8)
 $$
 One can check that the DT may be treated as similarity transformation 
 $$
 \rho[i+1]=
 \Big( 1+\frac{\mu_i-\nu_i}{\nu_i}P_i\Big)
 \rho[i]
 \Big(1+\frac{\nu_i-\mu_i}{\mu_i}P_i \Big)\eqno(6.9).
 $$

 Multiplying  (6.9)  by $P_{i+1}$ from the left side and by $P_i$ by 
the right side,
 changing $\rho[n+1] = \rho_n$ one obtains:
 $$
 P_{n+1}\rho_{n+1}P_n =P_{n+1}(1+ \frac{\mu_n-\nu_n}{\nu_n}P_n )\rho_nP_n(1+
 frac{\nu_n-\mu_n}{\nu_n}P_n )
 $$
 or, using the links (6.6) for the appropriate n,
 $$
 P_{n+1}(\nu_{n+1} A+z_{\nu,n+1})P_n = P_{n+1}(1+ \frac{\mu_n-\nu_n}{\nu_n}P_n )(\mu_n A P_n+
 z_{\mu,n} P_n)(1+frac{\nu_n-\mu_n}{\nu_n}P_n ).
 $$
 Simplifying, one arrives to the chain equation.
 $$
 (\nu_{n+1}-\nu_{n})P_{n+1} AP_n + (z_{\nu,n+1}-z_{\nu,n})P_{n+1}P_n = 0.  \eqno(6.10)
 $$
 or, similarly, after the right multiplication by $P_{n+1}$ and left one by 
$P_n$,  
 $$
(\mu_{n+1}-\mu_n) P_n AP_{n+1} + (z_{\mu,n+1}-z_{\mu,n})P_nP_{n+1}=0. \eqno(6.11)
 $$
 
 These equations are recurrences and define the solution via parameters. 
The dependence of the
 parameters on the additional 
variable t is introduced by the "master" equation (see \cite{LC}):
 $$
 \imath P_t = (AP-PAP)/\mu - (PA-PAP)/\nu.  \eqno(6.12)
 $$

The "minimal" closure $P_{n+1}= \lambda P_n=P_n\lambda$, 
is possible only if $\lambda=1$ for the $P_n$ is a projector. In this case  
 $$P_{n+1} = P_n = P, \qquad z_{\nu,n+1} = z_{\nu,n}= z_{\nu} $$ 
 So the equations (6.10,11) are valid authomatically.Other simple closure 
 $$
  P_{n+1} =a_nP_n+b_n
 $$
 with $a_n, b_n \in K$ leads to $P_{n+1} =P_n - (1-(-1)^n/2.$ 
 
 Next possibility is when some element A do not commute with $P_n$. For example 
 $$P_{n+1} = P_nA, \qquad z_{\mu,n+1}= \lambda_n z_{\mu} $$
 that for (6.11) leads to
 $$
 ( \lambda_n-1)z_{\mu,n}S_n^2=(\mu_n-\nu_n) S_n  
 $$
 where, $S_n=P_n A ,$ is proportional to some projector,
 or, by recurrence, it means
 $$
 P_n=p_0(t)A^n
 $$
 where the function $p_0(t)$ is defined from (6.12).
  
 Finally, we arrive to a conjecture:
 
 {\it The chain equations (6.11,12) may help to construct particular 
solutions of the compatibility 
 condition for the Lax pair (see \cite{LCUK}) if each $P_n$ 
is a solution of the equation (6.13).}
 
\section{\bf Conclusion}
 
We conclude that our
approach

a) give a connection between a class of a
bilinearizable NLPDE, and its Lax form. The class
is formed by a certain combination of binary 
Bell polynomials that satisfy the covariance
principle.

b) produce automatically a technique of
integration of the equations that belong to this
class. The underlying Darboux covariance of the
both linear operators  permits to generate a rich set of
solutions. Between them  are multisolitons,
positons, periodic and rational solutions
with abundant combinations that introduce their
interaction.[

We add that a platform for a nonabelian
generalization is prepared. 
The joint covariance principle application presumably cover the items of classification
that appears inside most of symmetry approaches \cite{MSY},\cite{LY} and contain the direct link
to solutions via dressing chains - iterations of DT.

Mention also the  generalization of the theory of 
small deformations of iterated transforms \cite{Le}  with respect to intermediate 
parameters that appear within the iteration procedure of bDT .   Infinitesimal invariance generated by extended Backlund 
transformation was considered in \cite{S} . The perturbation formulas allow to 
define and investigate generators of the corresponding group, that is a 
symmetry group of a given hierarchy associated with the ZS problem under 
consideration.

We continue to develop the dressing method for linear and soliton problems.
The procedure of dressing includes both solutions and potentials  of underlying
linear system. 
The other reductions and generalizations may be introduced if in the ring
there exist an involutive automorphism.

There are abundant possibilities for generalizations based on recent 
results of V.B. Matveev \cite{M2}, where the DT transformation operations
is defined via some automorphism at rings. 
\centerline{\bf Acknowledgments}
 
{99}


\begin{thebibliography}{99}
\bibitem{L} S. Leble  Darboux transforms algebras in 2+1 dimensions.
{\em Proceedings of NEEDS-91 Workshop, World Scientific, Singapore},
1991, p.53-61.
\bibitem{ZL} Zaitsev A.A., Leble S.B. Preprint
12.01.1999 math-ph/9903005; {\it Rep.on math. Phys.}, {\bf 46},165-174, (2000).
\bibitem{BZ} Borisov A, Zykov S. {\it Theor. Math. Phys.},{\bf 115.}, 199-214 (1998) 
\bibitem{M} Matveev V.B. Preprint LPTHE 79/06 (1979) 1-11; {\it Lett. in Math.
Phys.} {\bf 3} 217-219, (1979) .{\it Lett. in Math. Phys.}, {\bf3}, 503-512, (1979).
\bibitem{LZ} Leble S.B. Zaitsev A.A. 
Intertwine Operators and Elementary Darboux 
Transforms in Differential Rings and Modules. Preprint of Kaliningrad St. 
University, 20 December 1994. {\it Rep. Math. Phys.}, {\bf 39.}, 177-183 (1997).
\bibitem{U}  Ustinov N.V. {\it Darboux transformations for spectral problems 
with reductions}. 
Ph.D. theses, St.Petersburg, 1994.
\bibitem{NMPZ} Novikov S.P. Manakov S.V. Pitaevski L.P. Zakharov V.E.
{\it Theory of Solitons}. Plenum, New York, 1984.
 \bibitem{LU} Leble S., Ustinov N.: 
{\it Deep reductions for matrix Lax system,
invariant forms and elementary Darboux transforms. Proceeding of NEEDS-92
Workshop}, World Scientific, Singapore, 1993, p.34-41. {\it J. Phys. A: 
Math.Gen.}
{\bf26} (1993) 5007-5016.
\bibitem{Le} Leble S. {\it Computers Math.
Applic.}, {\bf 35}, 73-81, (1998).
\bibitem{Leb} Leble S. {\it Theor. Math. Phys.} {\bf 122}, 239-250, (2000)
\bibitem{Mih} Mikhailov A. 
%The reduction problem and the inverse scattering method. 
{\it Physica D} {\bf 3,} 
 73-117, (1981).
\bibitem{LC} Leble S.B. Czachor M. {\it Darboux-integrable nonlinear 
Liouville-von Neumann equation}
quant-ph/9804052, {\it Phys. Rev. E}, {\bf 58}, N6, (1998).
\bibitem{W}  J. Weiss, {\it J. Math. Phys},  {\bf 27},
  p.2647, (1986).
\bibitem{Sha}  A. Shabat 
{\bf Inverse Problems}, {\bf 8}, 303-308, (1992).
\bibitem{VSha}  A. Veselov, A.  Shabat, {\bf Funk. analiz i pril.} , {\bf 27}
  1, (1993)
  \bibitem{LY} Leznov A. Yuzbashyan E. 
  %Integrable mappings for noncommutative objects
 {\it Rep. Math. Phys.}, {\bf 43.}, 207-214, (1999).
 \bibitem{ZS} Zakharov V.E. Shabat A.B.  {\it Funk. analiz i pril.} , {\bf 8}
  p.43 (1974). 
 \bibitem{LCUK} Ustinov N. Leble S. Czachor M. Kuna M. 
{\it "Darboux-integration of $\imath\rho_t = 
 [H,f(\rho)]$"} quant-ph/0005030, submitted to Phys. Lett. A.
 \bibitem{CU} Czachor M. Ustinov N. {\bf New Class of Integrable
 Nonlinear von Neumann-Type equations}
 arXiv:nlinSI/0011013 7Nov 2000.
 \bibitem{CKLN} M. Czachor S.Leble M. Kuna and J.Naudts {\it Nonlinear von
 Neumann type equations. 
 "Trends in Quantum Mechanics" Proceedings of the International symposium, 
ed. H.-D. Doebner
et al} World Sci 2000, p 209-226.
 \bibitem{MS} Mikhailov A. Sokolov V. {\it Theor. Math. Phys.} , {\bf 122}, 
 72-83, (2000) 
  \bibitem{SR} R. Schimming, S. Z. Rida {\em Int. J. of Algebra and Computation} {\bf  6},
635--644 (1996).
\bibitem{Y} Yurov A. {\it "Darboux Transformations in Quantum Mechanics"}
. Kaliningrad State University,
Kaliningrad 1998.
\bibitem{GLNW} C. Gilson, F Lambert, J. Nimmo, R.
Willox {\em Proc. R. Soc. Lond A} {\bf  452},
223-234 (1996).
\bibitem{LS} F.Lambert and J.Springael 
%"Construction of B\"acklund transformations with binary Bell polynomials", 
{\it J.Phys.Soc.Jpn. \bf
66}, 2211--2213, (1997).
\bibitem{Sh} {\it "Dressing Chains and Lattices" Proceeding of the workshop
Nonlinearity, Integrability and all that: Twenty Years after NEEDS 79
ed M. Boiti et all}, World Scientific, Singapore, 2000, p.331-342.
\bibitem{M2}\bibitem{Mat} Matveev V B {\it Darboux Transformations in 
Associative Rings and
Functional-Diffrence Equations ed J Harnad and A Kasman 
"The Bispectral Problem"} (1998)
AMS series CRM PROCEEDINGS AND LECTURE NOTES v.14, p.211-226.  
\bibitem{MSY}  A. Mikhailov A.Shabat, R. Yamilov,
{\it Sov. Math. Doklady} {\bf 42}, 3-53 (1987).
\bibitem{S}Steudel H. Annalen der Physik
(Leipzig) {\bf 32} (1975) p.205,445,459].
\end{thebibliography}
\end{document}